\documentclass{article}

\usepackage{amssymb}
\usepackage{subcaption}
\usepackage{array}
\usepackage{tabulary,diagbox,pifont}
\usepackage[table,xcdraw]{xcolor}
\usepackage{spconf,amsmath,graphicx,url,bm,array,xcolor,booktabs, blindtext, enumitem, multirow}
\usepackage{footnote}
\usepackage[bottom]{footmisc}

\newcolumntype{C}[1]{>{\centering\arraybackslash}p{#1}}
\newlength\replength
\newcommand\repfrac{.33}

\newcommand\rulewidth{.6pt}
\newcommand\tdashfill[1][\repfrac]{\cleaders\hbox to \replength{%
  \smash{\rule[\arraystretch\ht\strutbox]{\repfrac\replength}{\rulewidth}}}\hfill}

\newcommand\tdotfill[1][\repfrac]{\cleaders\hbox to \replength{%
  \smash{\raisebox{\arraystretch\dimexpr\ht\strutbox-.1ex\relax}{.}}}\hfill}

\title{Multi-Channel Multi-Speaker ASR Using 3D Spatial Feature}
\name{Yiwen Shao$^{1\dagger*}$, Shi-Xiong Zhang$^{2\dagger}$, Dong Yu$^2$\thanks{$^*$This work was done while Yiwen was a research intern at Tencent AI Lab, USA. $\dagger$These authors contributed equally.} }
\address{$^1$Center for Language and Speech Processing, Johns Hopkins University, Baltimore, MD, USA\\
$^2$Tencent AI Lab, Bellevue, WA, USA}
\begin{document}
%
\maketitle
\begin{abstract}
Automatic speech recognition (ASR) of multi-channel multi-speaker overlapped speech remains one of the most challenging tasks to the speech community. In this paper, we look into this challenge by utilizing the location information of target speakers in the 3D space for the first time. To explore the strength of proposed the 3D spatial feature, two paradigms are investigated. 1) a pipelined system with a multi-channel speech separation module followed by the state-of-the-art single-channel ASR module; 2) a ``All-In-One" model where the 3D spatial feature is directly used as an input to ASR system without explicit separation modules. Both of them are fully differentiable and can be back-propagated end-to-end. We test them on simulated overlapped speech and real recordings. Experimental results show that 1) the proposed ALL-In-One model achieved a comparable error rate to the pipelined system while reducing the inference time by half; 2) the proposed 3D spatial feature significantly outperformed (31\% CERR) all previous works of using the 1D directional information in both paradigms.
\end{abstract}
%
\begin{keywords}
multi-channel ASR, multi-speaker ASR, audio-visual ASR, speech separation, 3D feature
\end{keywords}
\section{Introduction}
\label{sec:intro}
With the development of speech techniques and deep neural networks, dramatic improvement has been achieved on multiple automatic speech recognition (ASR) benchmarks \cite{hinton2012deep, wang2019espresso,shao2020pychain, zhang2020pushing}. However, it remains a challenging task for multi-channel multi-speaker overlapped speech recognition due to the interfering speakers or background noise \cite{subramanian2020far, gu2020multi, kanda2021investigation}.
To solve this problem, \cite{yu2020audio, yu2021audio, subramanian2021directional} adopt a pipeline-based paradigm where a preceding speech separation module separates clean target speech for the back-end standard ASR system to transcribe. These modules can be jointly trained with ASR criteria to achieve a boost on the final ASR performance.

To handle the multi-talker speech, the common approach is to use the spectrogram masking based separation \cite{wang2014training, wang2018supervised, hershey2016deep}, where the weight (mask) of the target speaker at each time-frequency (T-F) bin is estimated. To identify the target speech from the mixture,  additional information about the target speaker is required. Common clues including speaker's voiceprint based features \cite{vzmolikova2019speakerbeam,wang2018voicefilter}, vision clues \cite{gu2020multi,ephrat2018looking, wu2019time} and location based features \cite{chen2018multi,wang2018combining,gu2019neural} have been proven to be beneficial. This paper focuses on location based features due to its robustness and accuracy of visual face detection \cite{gu2020multi}.

However, there is one main issue with these location-based methods. To our best knowledge, all existing systems using spatial features only consider speech in a 1D scenario \cite{gu2020multi, subramanian2021directional, chen2018multi}. They assume that the sound sources traveled like a 2D planar wave and only the 1D directional of arrival (DoA) information (i.e. azimuth) is considered. As a result, when the coming directions of two simultaneous speech are close, even if they are far apart in distance, these systems will fail to separate them (see Fig.~\ref{fig:3D} (a)).

\noindent{\bf Contributions:} Previous works in this area are extended in two important directions. First, the traditional 1D directional features are generalized to 3D spatial features, which can leverage all the 3D location information of target speakers, including the azimuth, elevation, and distance respectively. Second, to further explore the strength of 3D spatial features and simplify the pipeline, we also design an All-In-One system that totally removes the separation module which can reduce training and inference time by more than a half, while still maintaining comparable results to the pipelined system. Experiments on both simulated and real data are reported. 


\begin{figure*}[t]
  \centering
  \includegraphics[width=\linewidth]{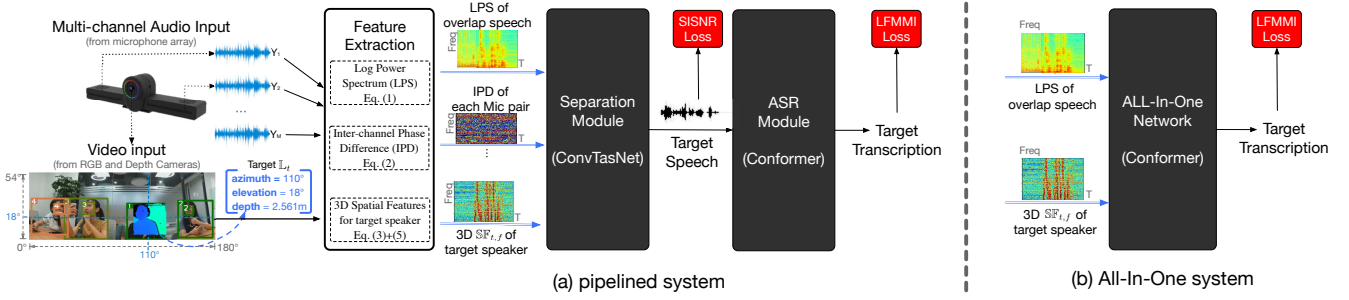}\vspace{-0.3cm}
  \caption{\small Two paradigms for multi-channel multi-speaker ASR. a) pipelined system with a multi-channel speech separation module followed by a single-channel ASR module; 2) ``All-In-One" model where 3D spatial features are directly used without separation module. Note the face and body tracking algorithm provides the location of each speaker, including the distance which is obtained from depth or binocular cameras.} \vspace{-0.3cm}
  \label{fig:diagram}
\end{figure*}


\vspace{-0.2cm}
\section{3D Spatial Feature}
\vspace{-0.1cm}
\label{sec:format}
\subsection{Spectral Feature}
The first step is to transform the $M$-channel noisy speech $y^{1:M}$ from time domain to frequency domain by short time Fourier transform (STFT). Given a window function $w$ with length
$N$, the complex spectrogram ${\bf Y}^m$ of the $m$-th channel can be calculated as: \begin{equation}
    {\bf Y}^m_{t,f} = \sum_{n=0}^{N-1}y^m[n]w[n-t]\cdot \text{exp}(-i\frac{2\pi n}{N}f)
    \label{stft}
\end{equation}
where $t$ and $f$ denote the frame index and frequency band index, respectively.
The logarithm power spectrum (LPS) of the reference
channel (the first channel in this work) is served as the spectral
feature, calculated by $\text{LPS}=\text{log}|{\bf Y}^{m=1}|^2$.
\subsection{Interchannel Phase Difference (IPD)}
IPD has been shown to be a strong cue as a spatial feature in many previous works \cite{chen2018multi, wang2018multi} for speech separation. We follow the practice in \cite{gu2020multi} of computing the phase difference between the selected $P$ pairs of microphones for both efficiency and better performance:
\vspace{-0.1cm}
\begin{equation}
     \text{IPD}^{(m)}_{t,f} = \angle {\bf Y}^{m_1}_{t,f} - \angle {\bf Y}^{m_2}_{t,f}
    \label{ipd}
\end{equation}
where $\angle {\bf Y}^{m_1}_{t,f}$ is the angle of this complex number and $(m)$ is the index for
mic-pair $(m_1, m_2)$. These are concatenated to form the IPD
features: $\textbf{IPD} = [..., \text{IPD}^{(m)}_{t,f}
, ..., ]_{P\times T\times F}$ .
\vspace{-0.2cm}

\subsection{3D Spatial Feature}
Spatial Feature (SF) or angle feature (AF) previously proposed in \cite{gu2019neural}, is used to indicate the dominance of the target source in the T-F bins. We define a target-dependent phase difference (TPD) as a phase delay caused by a theoretical wave (with frequency $f$) that travels from target location $\mathbb L_t$, measured at the $m$-th microphone pair at time $t$  \cite{gu2020multi}. Generally, the spatial features of a target speaker are formulated as the cosine distance between the target-dependent phase difference (TPD) and the interchannel phase differences (IPD):
\vspace{-0.2cm}
\begin{equation}  
    \mathbb{SF}_{t,f} =\sum_{p=1}^P\langle {\bm e}^{\text{TPD}_{t,f}^{(m)}(\mathbb L)}, {\bm e}^{\text{IPD}_{t,f}^{(m)}}\rangle 
    \label{SF} \vspace{-0.1cm}
\end{equation}
where ${\bm e}^{(\cdot)}=\begin{bmatrix} \cos(\cdot) \\ \sin(\cdot)\end{bmatrix}$ and $\text{TPD}_{t,f}^{(m)} (\mathbb L)$ depends on target speaker's location $\mathbb L$ obtained from the vision. 
It measures the phase difference at mic-pair $(m)$ if there is a wave spread from location $\mathbb L$. 
Intuitively, if the speaker at location $\mathbb L$ is speaking, this \emph{theoretical} phase difference $\text{TPD}_{t,f}^{(m)} (\mathbb L)$ from vision will be very similar to the \emph{observed} phase difference $\text{IPD}_{t,f}^{(m)}$. Thus $\mathbb{SF}_{t,f}$ will be close to 1, otherwise close to 0.


\begin{figure}[htb]
    \centering
    \includegraphics[width=9cm]{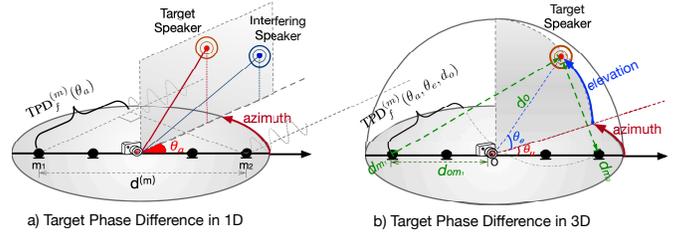}
    \vspace{-0.5cm}
    \caption{\small The illustration of 1D and 3D scenarios. Fig~\ref{fig:3D} a) shows inseparable issue when azimuths of two simultaneous speech are close.} \vspace{-0.3cm}
    \label{fig:3D}
\end{figure}

\vspace{-0.1cm}
\subsubsection{Target Phase Difference in 1D}
In 1D cases, the planar wave propagation model was assumed and only the directional information--azimuth $\theta_a$ of the target speaker is known. Thus, the location $\mathbb{L}=[\theta_{a}]$ and the target phase difference $\text{TPD}_{t,f}^{(m)} (\mathbb L)$ can be obtained by: \vspace{-0.1cm}

\begin{equation}
    \text{TPD}_{t,f}^{(m)} (\theta_a) = \frac{2\pi f}{c(F-1)} \cdot f_s \cdot d_{(m)} \text{ cos }\theta_a  
    \label{1dtpd} \vspace{-0.1cm}
\end{equation}
where $d^{(m)}$ is the distance between the microphone-pair $(m)$, $f_s$ is the sampling rate, $c$ is the sound velocity and $F$ is the total number of frequency bands. 

\vspace{-0.1cm}
\subsubsection{Target Phase Difference in 3D}
\vspace{-0.1cm}

One problem of the $\mathbb{SF}_{t,f}$ in 1D is its limited discriminative power when any of the interference speakers shares close azimuth with the target speaker, even they are far apart. As we observed in the empirical study \cite{chen2018multi, gu2019neural}, when the speakers are close, e.g., the azimuth difference (AD) between speakers is less than 15 degrees, the spatial information calculated from azimuth is not sufficient to distinguish between speakers. Thus, it is necessary to generalize the TPD to 3D space with full location information $\mathbb L=[\theta_a, \theta_e, d_o]$. We illustrate it in Fig.~\ref{fig:3D} and the generalized $\text{TPD}_{t,f}^{(m)} (\mathbb L)$ in 3D becomes
\vspace{-0.1cm}
\begin{equation}
\begin{split}
   & \text{TPD}_{t,f}^{(m)} (\theta_a, \theta_e, d_o) = \frac{2\pi f}{c(F-1)} \cdot f_s \cdot (d_{m_1} - d_{m_2}) \\
   &     d_{m_i} = \sqrt{d^2_{om_i}+d^2_o-2d_{om_i}d \cos \theta_a \cos \theta_e} ~~_{\forall i \in \{1,2\}}
\end{split}
\label{eq:TPD3D} \vspace{-0.1cm}
\end{equation}
where $\theta_a$ is azimuth, $\theta_e$ is elevation. $d_o$, $d_{m_1}$ and $d_{m_2}$ are distances between the target speaker and the camera, the $m_1$-th and the $m_2$-th microphone, respectively. $d_{om_i}$ is the distance between the camera and the $m_i$-th microphone (as shown in Fig.~\ref{fig:3D}). Here the camera is at the center of microphone array.

\begin{figure}\vspace{-0.1cm}
    \centering
    \begin{minipage}[b]{0.3\linewidth}
    \includegraphics[width=25mm,scale=1]{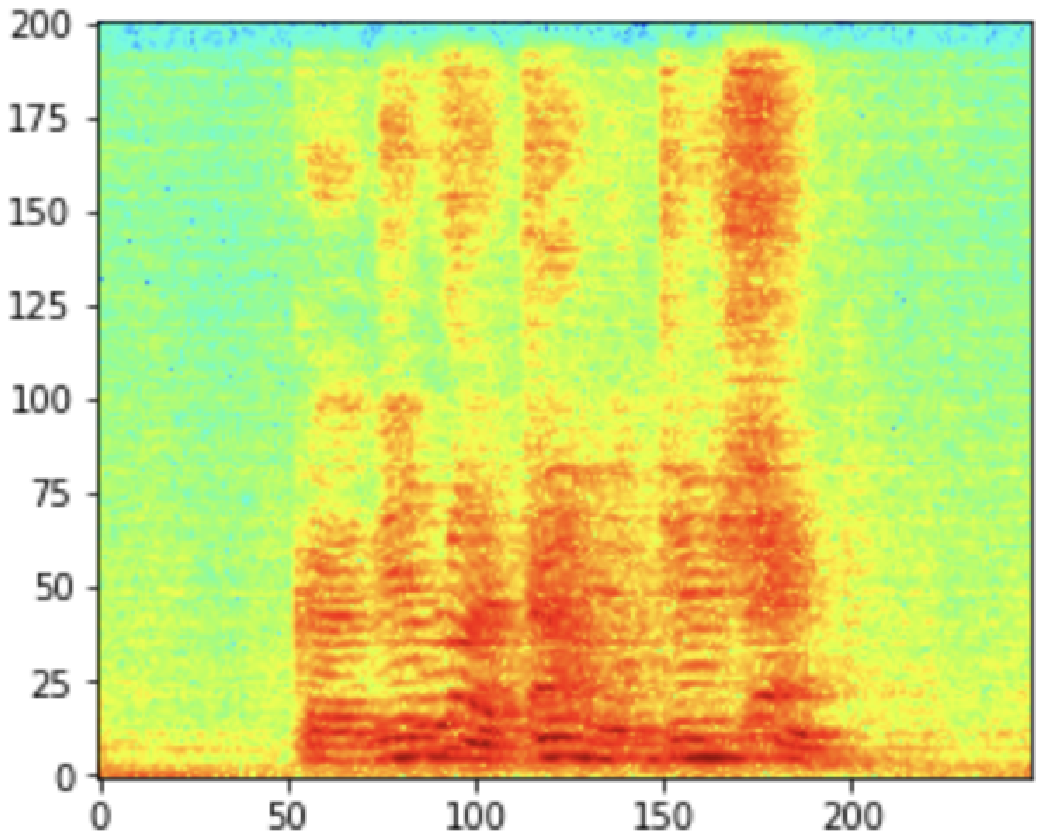}
    \subcaption{Overlap speech}
    \end{minipage}
    \begin{minipage}[b]{0.3\linewidth}
    \includegraphics[width=25mm,scale=1]{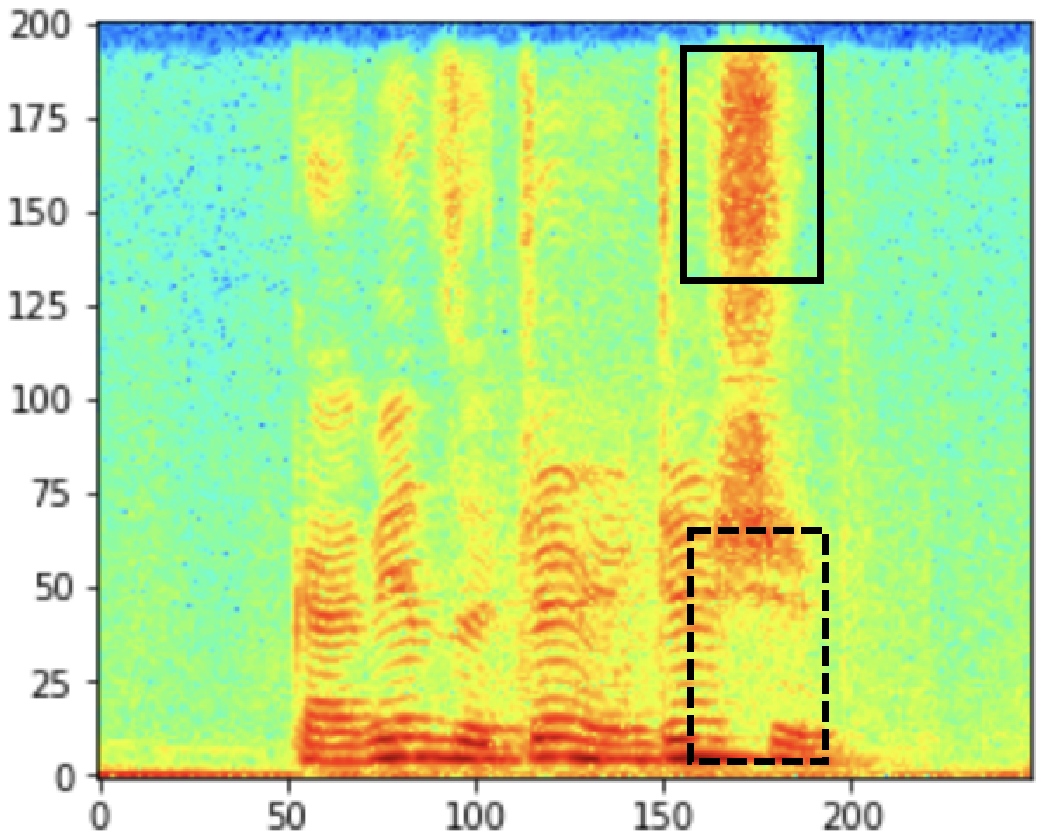}
    \subcaption{Target}
    \end{minipage}
    \begin{minipage}[b]{0.3\linewidth}
    \includegraphics[width=25mm,scale=1]{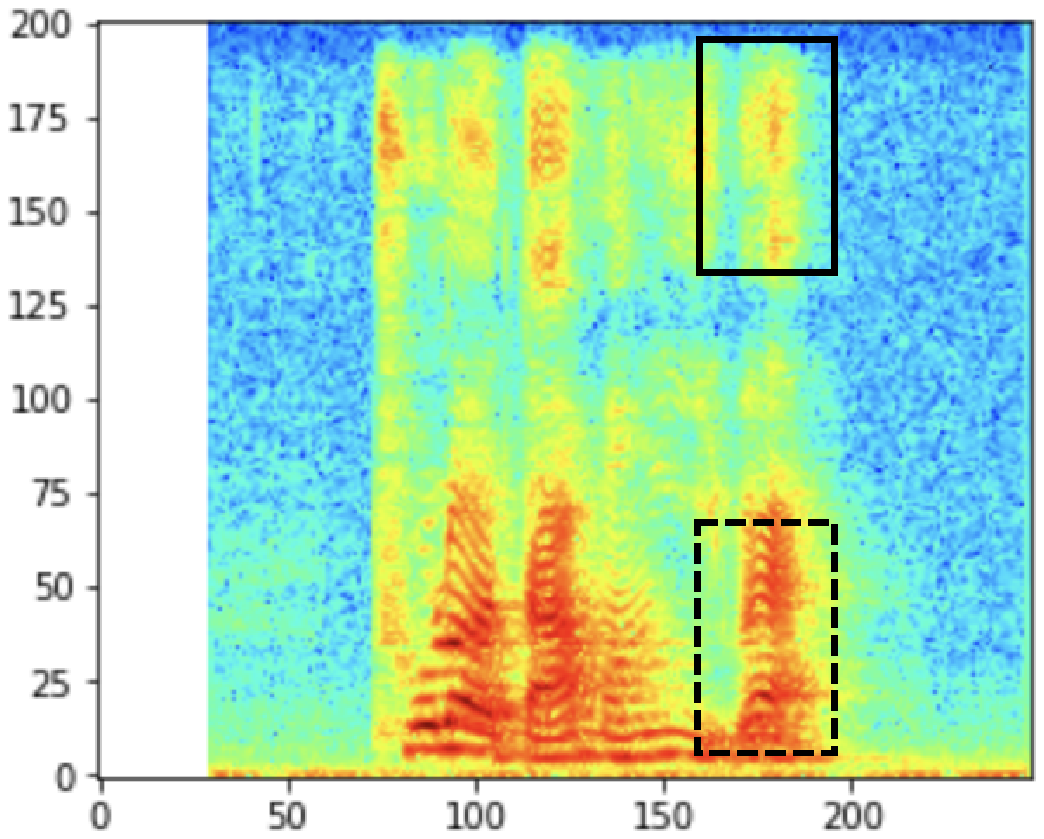}
    \subcaption{Interference}
    \end{minipage}
     \begin{minipage}[b]{0.3\linewidth}
    \includegraphics[width=25mm,scale=1.0]{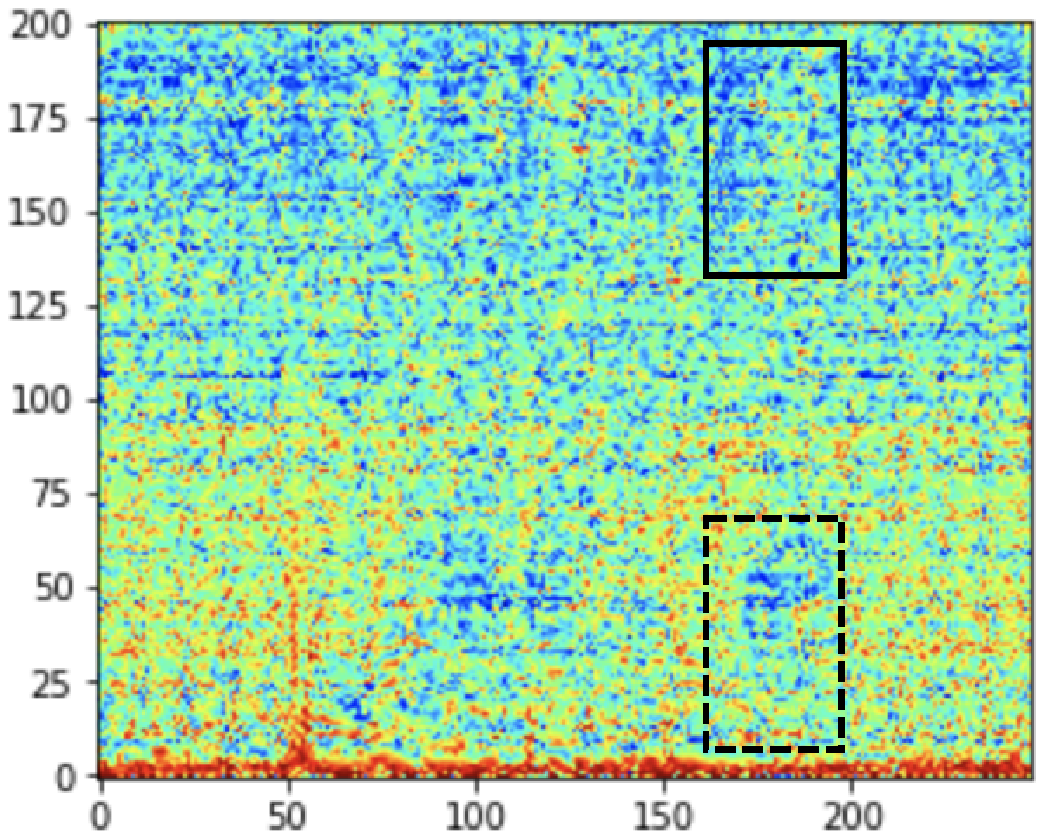}
    \subcaption{1D $\mathbb{SF}_{t,f}$}
    \end{minipage}
     \begin{minipage}[b]{0.3\linewidth}
    \includegraphics[width=25mm,scale=1.0]{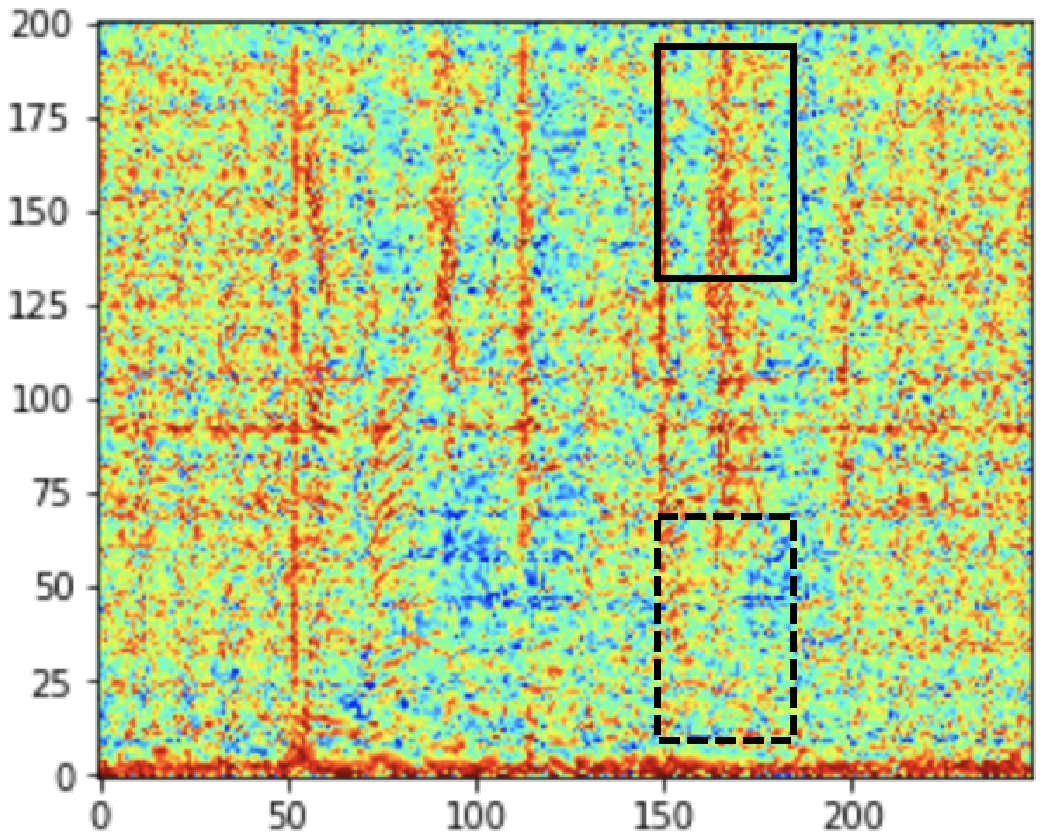}
    \subcaption{3D $\mathbb{SF}_{t,f}$}
    \end{minipage} \vspace{-0.3cm}
    \caption{\small An illustration of spatial features in 1D and 3D. Black solid rectangle marks the T-F bins dominated by target speaker while the dashed rectangle marks T-F bins dominated by interfering speaker. }\vspace{-0.4cm}
    \label{fig:spec}
\end{figure}

Fig.~\ref{fig:spec} is an example of a mixed speech of two. Comparing Fig.~3 (d) and (e) shows that the proposed 3D $\mathbb{SF}_{t,f}$ not only exhibits the dominance of speakers in T-F bins precisely, but also possesses stronger contrast than its 1D counterpart.  \vspace{-0.2cm}

\section{Systems for multi-speaker ASR}
\vspace{-0.1cm}

\subsection{Pipelined System}
\vspace{-0.1cm}

There are two paradigms for multi-speaker ASR --- Pipelined approaches and All-In-One approaches. 
As shown in Fig.~\ref{fig:diagram}, the pipelined system takes multi-channel $Y^{1:M}$ along with target's location information $\mathbb{L}=(\theta_a, \theta_e, d_o)$ as input, followed by a single-channel ASR module.

\vspace{-0.2cm}
\subsubsection{Separation Module}
\vspace{-0.2cm}
\textbf{Features:} For short time Fourier transform (STFT) setting, we use 32ms sqrt hann window and 16ms hop size at a sampling rate of 16kHz, resulting in a frame length of 512 and hop size of 256. 512-point FFT is used to extract 257-dimensional LPS. 6 pairs of microphones are selected for IPD and TPD computation, which are $(0, 7), (1, 6), (2, 5), (3, 4), (4, 7), (3, 4)$. The total dimension of the input feature after concatenation is $257\times (1+6+1) = 2056$.

\noindent\textbf{Architecture:} 
A Conv-TasNet is used for speech separation with the same settings as \cite{gu2020multi}. It consists of multiple stacked dilated 1-dimensional convolution blocks to increase the receptive field over the input sequence. We use batch normalization to reduce the training and evaluation duration. 
An iSTFT convolution 1d layer is placed at the end of the model to convert the estimated target speaker complex spectrogram back to the single-channel waveform. 

\noindent\textbf{Pretraining:} We pretrain the separation module on the simulated data with scale-invariant signal to noise ratio (Si-SNR) as a criterion and the reverberant clean speech as supervision. The signal-to-distortion ratio (SDR) and perceptual evaluation of speech quality (PESQ) scores computed with the reverberant clean speech as the reference are used as the metric for speech separation. Our best pretrained separation module achieves 12.8 dB of SDR and 3.35 of PESQ on the test set.

\vspace{-0.2cm}
\subsubsection{ASR Module}\vspace{-0.2cm}
\textbf{Features:} Log Mel-Filterbank (LFB) acoustic features with 40 bins are
used, which are extracted with a 25ms window, 10ms hop-length
at a sample rate of 16kHz.

\noindent\textbf{Architecture: }For better performance on overlapped speech as shown in \cite{yu2020audio, yu2021audio}, Lattice-free MMI (LFMMI) loss in k2 \cite{k2} is used as the training criterion. As to model structure, two convolution sub-sampling layers with kernel size 3*3 and stride 2 are used in the front of the 12-layer 4-head Conformer encoder \cite{gulati2020conformer}. Each
conformer layer uses 384 attention dimensions and 2048 feed-forward dimensions. 

\noindent\textbf{Pretraining:} We pretrain the ASR module on either dry clean (without reverberation) or reverberant clean data. Adam optimizer is used with a learning rate schedule with 5,000 warm-up steps. Table \ref{tab:my-table} shows the character error rates (CERs) we get from different training set. Note that "separated" represents the output from the best pretrained separation module we trained. Since we are not doing any dereverberation in this system, separated speech is also reverberant. In this case, the best CER on reverberant clean test data (8.66\%) can be treated as the upper bound of the ASR module when given perfect separated speech. And the best CER on separated speech (16.41\%) can be interpreted as the lower bound, which shows how the pipelined system performs when we simply connect two pretrained modules. 

\vspace{-0.2cm}
\subsubsection{Joint Training}\vspace{-0.2cm}
It is exhibited in previous works that end-to-end joint training of separation module and ASR module using ASR criterion (LFMMI loss in our case) can bring further improvement to the pipelined system \cite{yu2020audio, yu2021audio, chang2019mimo}. We compare several different training strategies as shown in Table \ref{tab:my-table}. We notice that when there is no pretraining (System 1), the pipelined system fails to converge. The best practice is to start from pretrained models, and do joint finetuning on both modules using mixed speech (System $4$). It gives the best CER of 10.50\% on simulated test and 5.03\% on the real recordings, which is close to the upper bound of 8.66\% in Table \ref{tab:table1}.

\begin{table*}[t]
\setlength{\tabcolsep}{4pt}
\resizebox{\linewidth}{!}{
\begin{tabular}{|l||ccc|ccc||c||ccc||c||c|c|}
\hline
\multicolumn{1}{|c||}{sys}                      & \multicolumn{6}{l||}{\cellcolor[HTML]{EFEFEF}\begin{tabular}[c]{@{}c@{}} \textbf{Pipelined System}\\ Separation module (SISNR loss)$\rightarrow | \rightarrow$ASR module (LFMMI loss)\end{tabular}}                                                                                                                                                & {\cellcolor[HTML]{EFEFEF}\begin{tabular}[c]{@{}c@{}} \textbf{All-in-One System}  \\ (LFMMI loss) \end{tabular}}                         & \multicolumn{1}{c}{}                                                                         & \multicolumn{1}{c}{}                                                                                   & \multicolumn{1}{c||}{}                                                                        & \multicolumn{1}{c||}{SDR of}                                                                                              & \multicolumn{2}{c|}{CER(\%)}      \\ \cline{2-8} \cline{13-14} 
\multicolumn{1}{|c||}{ID} & Inputs               & \begin{tabular}[c]{@{}c@{}}pretrain\\ (overlap)\end{tabular} & \begin{tabular}[c]{@{}c@{}}joint \\ finetune\end{tabular} & Inputs               & \begin{tabular}[c]{@{}c@{}}pretrain\\ (nonoverlap)\end{tabular} & \begin{tabular}[c]{@{}c@{}}joint \\ finetune\end{tabular} & \begin{tabular}[c]{@{}c@{}}Inputs\\ \end{tabular} & \multicolumn{1}{c}{\multirow{-3}{*}{\begin{tabular}[c]{@{}c@{}}\# param\\ (M)\end{tabular}}} & \multicolumn{1}{c}{\multirow{-3}{*}{\begin{tabular}[c]{@{}c@{}}Training\\ time/iter \\ (s)\end{tabular}}} & \multicolumn{1}{c||}{\multirow{-3}{*}{\begin{tabular}[c]{@{}c@{}}Decoding\\ RTF\end{tabular}}} & \multicolumn{1}{c||}{\multirow{-2}{*}{\begin{tabular}[c]{@{}c@{}}Separation \\ module \end{tabular}}} & simulated          & real         \\ \hline  \hline
1                                           & LPS+IPD+3D           & \ding{55}                                                        & \checkmark                                                              & LFB                  & \ding{55}                                                                & \checkmark                                                              &                                                                & 45.4                                                                                         & 2.35                                                                                                   & 0.0264                                                                                        &     --                                                                                                               & \multicolumn{2}{c|}{not converge} \\ 
2                                           & LPS+IPD+3D           & \checkmark                                                        & \ding{55}                                                                    & LFB                  & \checkmark                                                             & \ding{55}                                                                    &                                                                & 45.4                                                                                         & 2.35                                                                                                   & 0.0264                                                                                        & 12.8 dB                                                                                                              & 16.41              &    8.03          \\ 
\textbf{3}                                           & LPS+IPD+3D           & \checkmark                                                        & \ding{55}                                                                    & LFB                  & \checkmark                                                             & \checkmark                                                              &                                                                & 45.4                                                                                         & 2.35                                                                                                   & 0.0264                                                                                        & 12.8 dB                                                                                                               & \textbf{14.65}              & \textbf{7.08}         \\
\textbf{4}                                           & LPS+IPD+3D           & \checkmark                                                        & \checkmark                                                              & LFB                  & \checkmark                                                             & \checkmark                                                              &             \multirow{-2}{*}{N/A}                                & 45.4                                                                                         & 2.35                                                                            & 0.0264                                                                                        & 11.0 dB                                                                                                              & \textbf{10.50}              & \textbf{5.03}         \\
5                                           & LPS+IPD+1D           & \checkmark                                                        & \checkmark                                                              & LFB                  & \checkmark                                                             & \checkmark                                                              &                                                                & 45.4                                                                                         & 2.31                                                                                                   & 0.0261                                                                                        & ~~9.7 dB                                                                                                               & 17.16              &    9.15         \\
6                                           & LFB+3D               & \checkmark                                                        & \checkmark                                                              & LFB                  & \checkmark                                                             & \checkmark                                                              & \multirow{-6}{*}{}                                             & 44.3                                                                                         & 2.29                                                                                                   & 0.0252                                                                                        & 10.0 dB                                                                                                             & 15.03              &   7.74          \\ \hline  \hline
7                                           & \multicolumn{1}{l}{} & \multicolumn{1}{l}{}                                       & \multicolumn{1}{l}{}                                            & \multicolumn{1}{l}{} & \multicolumn{1}{l}{}                                            & \multicolumn{1}{l||}{}                                            & \multicolumn{1}{l||}{LPS+IPD+3D}                                & 37.5                                                                                         & 0.92                                                                                                   & 0.0157                                                                                        &                                                                                                                    & 18.24              &   8.16           \\
8                                           & \multicolumn{1}{l}{} & \multicolumn{1}{l}{}                                       & \multicolumn{1}{l}{}                                            & \multicolumn{1}{l}{} & \multicolumn{1}{l}{}                                            & \multicolumn{1}{l||}{}                                            & \multicolumn{1}{l||}{LFB+IPD+3D}                                & 37.2                                                                                         & 0.91                                                                                                   & 0.0157                                                                                        &                                                                                                                    & 17.51              &   7.85           \\
\textbf{9}                                           & \multicolumn{6}{c||}{\multirow{-2}{*}{N/A}}                         & \multicolumn{1}{l||}{LFB+3D}                                    & 36.4                                                                                         & 0.78                                                                                                   & 0.0121                       &    {\multirow{-2}{*}{N/A}}                                                                                                                 & \textbf{14.37}              & \textbf{6.35}         \\
10                                          & \multicolumn{1}{l}{} & \multicolumn{1}{l}{}                                       & \multicolumn{1}{l}{}                                            & \multicolumn{1}{l}{} & \multicolumn{1}{l}{}                                            & \multicolumn{1}{l||}{}                                            & \multicolumn{1}{l||}{LFB+1D}                                    & 36.4                                                                                         & 0.78                                                                                                   & 0.0121                                                                                        &                                                                                                                    & 22.75              & 9.22         \\ \hline
\end{tabular}
}
\captionsetup{font=small}
\caption{\small Comparing pipelined systems and All-In-One systems, in terms of decoding speed and the CER (\%) on both Aishell-1-mixed simulated test and real recordings. The joint finetuning of the pipelined system uses LFMMI criteria for both separation and ASR modules. Decoding RTF was tested on the GPU (Nvidia V100, batch size=1) with perfetched input and no model quantization. Note, the all-in-one system 9 performs better than the pipelined system 3 in CER and with only half the decoding time. Another advantage of the system 9 is it uses LFB and 3D spatial features only, which makes the all-in-one model independent to microphone arrays typologies.} \vspace{-0.3cm}
\label{tab:my-table}
\end{table*}


\vspace{-0.2cm}
\subsection{All-In-One System}
\vspace{-0.1cm}
\textbf{Motivation:} Although pipelined systems achieve great results with 3D spatial feature, there still exists several limitations: 1) Requirement of parallel reverberant clean data, 2) mismatch of training scale between the separation module (chunk) and ASR module (utterance) and c) lack of usage of contextual and semantic information. For these reasons, we believe it is beneficial to simplify the pipelined system into a "All-In-One" model that removes the separation module.

\noindent\textbf{Architecture: }In the All-In-One system, to focus on the 3D spatial feature we propose, we further reduce the input features to only contain $\mathbb {SF}$ and 40-dimension LFB.  
We unify the window length and hop size to be 25ms and 10ms, resulting in a 400-point FFT and 201 dimensions for $\mathbb{SF}$. We simply concatenate 2 features and the final input feature dimension is $40+201=241$. One advantage of this setup is all the spatial and multi-channel information have been baked into $\mathbb {SF}$, thus the model becomes independent to the microphone array typologies. The model structure is the same 12-layer 4-head conformer in the ASR module described above.


\begin{savenotes}
\begin{table}[b]\vspace{-0.2cm}
\small
\setlength{\tabcolsep}{5pt}
\begin{tabular}{l|l|l|l|l|}
\cline{1-5}
\multicolumn{1}{|l|}{\diagbox{Train}{CER(\%)}{Test}}                  & \begin{tabular}[c]{@{}l@{}}dry\\ clean\end{tabular} & \begin{tabular}[c]{@{}l@{}}reverberant\\ clean\end{tabular} & \begin{tabular}[c]{@{}l@{}}overlapped\\ speech\end{tabular} & \begin{tabular}[c]{@{}l@{}}separated\\ speech\end{tabular} \\ \hline
\multicolumn{1}{|l|}{dry clean}         & 7.00\footnote{We achieved CER=6.3\% with model averaging \cite{izmailov2018averaging} during decoding which is comparable to the state-of-the-art results in \cite{watanabe2018espnet}. For simplicity we did not apply model averages for the rest of experiments.}                                                & 13.05                                                       & 101.03                                                 & 26.92                                                      \\ \hline
\multicolumn{1}{|l|}{reverberant clean} & 7.06                                                & \textbf{8.66}                                                       & 100.54                                                 & \textbf{16.41}                                                     \\ \hline
\end{tabular} \vspace{-0.2cm}
 \caption{\small CERs (\%) on different Aishell-1 test set trained with dry/reverberant clean speech.}
 \label{tab:table1} 
\end{table}
\end{savenotes}

\vspace{-0.2cm}
\section{Experiments}
\vspace{-0.2cm}
As there is no existing dataset containing overlapped speech with 3D location information, we prepare the training data by simulation. The open-source Chinese Mandarin speech corpus AISHELL-1 \cite{bu2017aishell} is selected as the clean data, which contains a 150-hr training set, a 10-hr development set, and a 5-hr test set. For simulated data, 100,000 sets of room impose response (RIRs) are generated. 
The room size is ranging from [3, 3, 3] to [10, 8, 5] meters. 
An 8-element non-uniform linear array with spacing 15-10-5-20-5-10-15 cm is used as the recorded sensors. The multi-channel signals are generated using the image-source method (ISM) \cite{lehmann2008prediction}. The reverberation time T60 is ranging 0.05s to 0.7 seconds. The signal-to-interference rate (SIR) is ranging from -6 to 6 dB. The sampling rate is 16kHz. Each utterance is simulated 3 times, mixed with random interfering utterances from the same dataset split. 

To evaluate the proposed 3D spatial features in real-life applications, the systems were also tested on our in-car data which contains 2000 utterances recorded using our device in daily car scenarios and $10000$-hr simulated data is used to train the in-car models. The target speech of in-car tests is the general voice commands of controlling the car, such as "turn on the air conditioner" or "open the back sit window". The interference speech is replayed from loudspeakers. There are 4 potential speakers in the car: the main driver, the co-driver and two passengers sitting in the back. The main driver's voice is taken as the target. Note the azimuths of the main driver and the main-back passenger (loudspeaker) are very close, though their elevations and distances are far apart. 
In our experiments on simulated data, we use the oracle location information in both training and test time. On car datasets, this information can be obtained by face detection and tracking techniques with depth cameras \cite{king2009dlib}.

A complete comparison between the pipelined systems and the All-In-One system is shown in Table 2. Systems 1-6 are pipelined systems and Systems 7-10 are All-In-One systems. When compared with the pipelined systems, although the All-In-One systems still fall behind in CER, the training time and decoding time is speedup by $3\times$ and $2\times$ respectively. In addition, All-In-One systems are trained from scratch which further reduces the effort on pretraining. Thus we believe when empowered by well-designed features like 3D spatial features, the All-In-One system has good potential and is a promising direction for further study.

\vspace{-0.3cm}
\section{Conclusions}
\vspace{-0.2cm}
In this work, we study the application of 3D spatial features for multi-channel multi-speaker overlapped speech recognition. It shows great improvement over the previous 1D feature and achieves 31+\% relative CERR on both simulated and real data. We also explore the idea of the All-In-One model with the proposed 3D feature as input and obtain comparable results. We believe when paid more effort and attention, the All-In-One model has the potential to surpass the prevalent pipelined systems in terms of both efficiency and accuracy.



\vfill\pagebreak

\footnotesize{
\bibliographystyle{IEEEbib}
\bibliography{main}
}

\end{document}